# Mechanochemical reaction in graphane under uniaxial tension

N.A.Popova and E.F.Sheka

*Peoples' Friendship University of the Russian Federation, 117198 Moscow, Russia*

**Abstract**. The quantum-mechanochemical-reaction-coordinate simulations have been performed to investigate the mechanical properties of hydrogen functionalized graphene. The simulations disclosed atomically matched peculiarities that accompany the deformation-failure-rupture process occurred in the body. A comparative study of the deformation peculiarities related to equi-carbon-core (5,5) nanographene and nanographane sheets exhibited a high stiffness of both bodies that is provided by the related hexagon units, namely benzenoid and cyclohexanoid, respectively. The two units are characterized by anisotropy in the microscopic behavior under elongation along mechanochemical internal coordinates when the later are oriented either along (*zg*) or normally (*ach*) to the C-C bonds chain. The unit feature in combination with different configuration of their packing with respect to the body C-C bond chains forms the ground for the structure-sensitive mechanical behavior that is different for *zg* and *ach* deformation modes. Hydrogenation of graphene drastically influences behavior and numerical characteristics of the body making tricotage-like pattern of the graphene failure less pronounced and inverting it from the *zg* to *ach* mode as well as providing less mechanical resistance of graphane it total.

## 1. Introduction

Among chemically modified graphenes, graphane takes a peculiar position due to optimistic expectations of its production in valuable quantities, on one hand, and intense studies of its electronic, vibrational, and mechanical properties, on the other. Among the studies, dominate extended theoretical calculations while experimental investigations have been restricted until now by a single work [1]. What makes graphane so attractive? Sure, this firstly concerns electronic properties of a new 2D solid provided by opening bad gap. Secondly, graphane is expected to possess unique mechanical strength comparable with that one of graphene, which is supported by numerous calculations [2-7]. The latter, related to the main topic of the current paper, were mainly concentrated on the evaluation of elastic parameters of the 2D graphane crystal within the framework of an isotropic continuum elastic shell model and only Ref. 2 and 7 raised the issue on spatial dependence of mechanical behavior of graphane.

Calculations [3-7] were performed by using DFT unit-cell approximation with periodic boundary conditions (PBC). All of them are mutually consistent concerning elastic parameters obtained but none of them discloses how the graphane deformation proceeds and what mechanism of the graphane body fracture takes place. In contrast, a general view on the graphane fracture behavior has been presented by molecular dynamics calculations of a large graphane sheet [2]. The computations have disclosed a crack origin in the sheet body, the crack gradual growth, and a final sheet fracture. The situation is similar to that one occurred by the middle of 2009 concerning deformation of graphene [8]. A large number of computations, preferably in supercell-PBC DFT, provided elastic parameters of graphene when the mechanical behavior of the graphene nanostructures remained unknown. Molecular dynamic calculations applied to long armchair-edged graphene nanoribbon [8] have thrown light on a peculiarity of the ribbon deformation disclosing the formation of one-atom chain in due course of uniaxial tension, which was confirmed experimentally [9, 10]. This peculiar behavior has been highlighted and given in detail later from the viewpoint of mechanochemical reaction that lays the foundation of the deformation-failure-rupture process occurred in nanographenes [11, 12]. The quantum-mechanochemical-reaction-coordinate (QMRC) approach used in the study has disclosed atomically matched peculiarities of the mechanical behavior, highlighted the origin of the mechanical anisotropy of graphene, and made allowance for tracing a deformation-stimulated change in the chemical reactivity of both nanographene body and its individual atoms. And once again, the molecular dynamics study has shown that the fracture of not already graphene but graphane occurs in a peculiar manner pointing that more detail molecular theory insight is needed.

In this paper, we present a comprehensive study of the mechanical properties of hydrogen functionalized graphene using QMRC simulations. We have found that hydrogenation has significant influence on the mechanical properties. Our simulations reveal that the tensile strength, fracture strain, and Young's modulus are sensitive to the functionalization. Basing on a detailed analysis of the deformation features, we attribute these significant changes in mechanical properties to the substitution of the benzenoid units of graphene by cyclohexanoids of graphane. The simulations reveal that the elastic part of the deformation is vibration-involving so that the ratio of Young's modules of graphene and graphane are in a good consistence with the ratio of squared frequencies of the species valence C-C vibrations and/or internal phonons.

## 2. Methodology and computational details

As mentioned in Introduction, graphane is the subject of computational studies mainly. Through over the computations, it is considered as a product of the graphene hydrogenation that preserves a honeycomb composition of the carbon core. However, in contrast to unified benzenoid structure of graphene units, the honeycomb cell of the 2D hydrogenated graphene crystal should be attributed to a particular cyclohexanoid unit, whose rich conformer biography, in contrast to benzenoid, has to be taken into account. This circumstance raises a definite question: which conformational honeycomb structure of hydrogenated graphene should be assigned as graphane? Is it a strictly determined monoconformer structure or is it just any arbitrary hydrogenated graphene that is implied in a predominant majority of the computational studies cited above? Historically, a polyhydride of graphene $(CH)_n$ named graphane was introduced as a regular package of chairlike cyclohexanoid units [13]. This assignment was supported in [1] attributing the product of both-side hydrogenated membrane fixed by perimeter to this very structure. In spite of no evidence of regular 2D packing of other cyclohexanoid conformers has been so far obtained, computational studies do not exclude 2D $(CH)_n$ crystals consisting of boat- [13], table- [14], zigzag- and/or armchair- [15], and stirruplike [16] cyclohexanoids thus making graphane's isomerism quite rich. A comprehensive study of stepwise hydrogenation of graphene [17] has shown that regularly packed chairlike cyclohexanoids provide not only energetically, but also topologically the most preferable graphane isomorph. This very isomorph will be considered in the current paper.

The studied nanographane sheet is shown in Fig.1. The species was obtained in due course of stepwise hydrogenation of (5,5) nanographene [17] that was, in its turn, subjected to uniaxial tensile deformation as well [11, 12]. The identity of the carbon cores of both objects provides a perfect basis for a comparative study of the mechanical behavior of the pristine and chemically modified bodies.

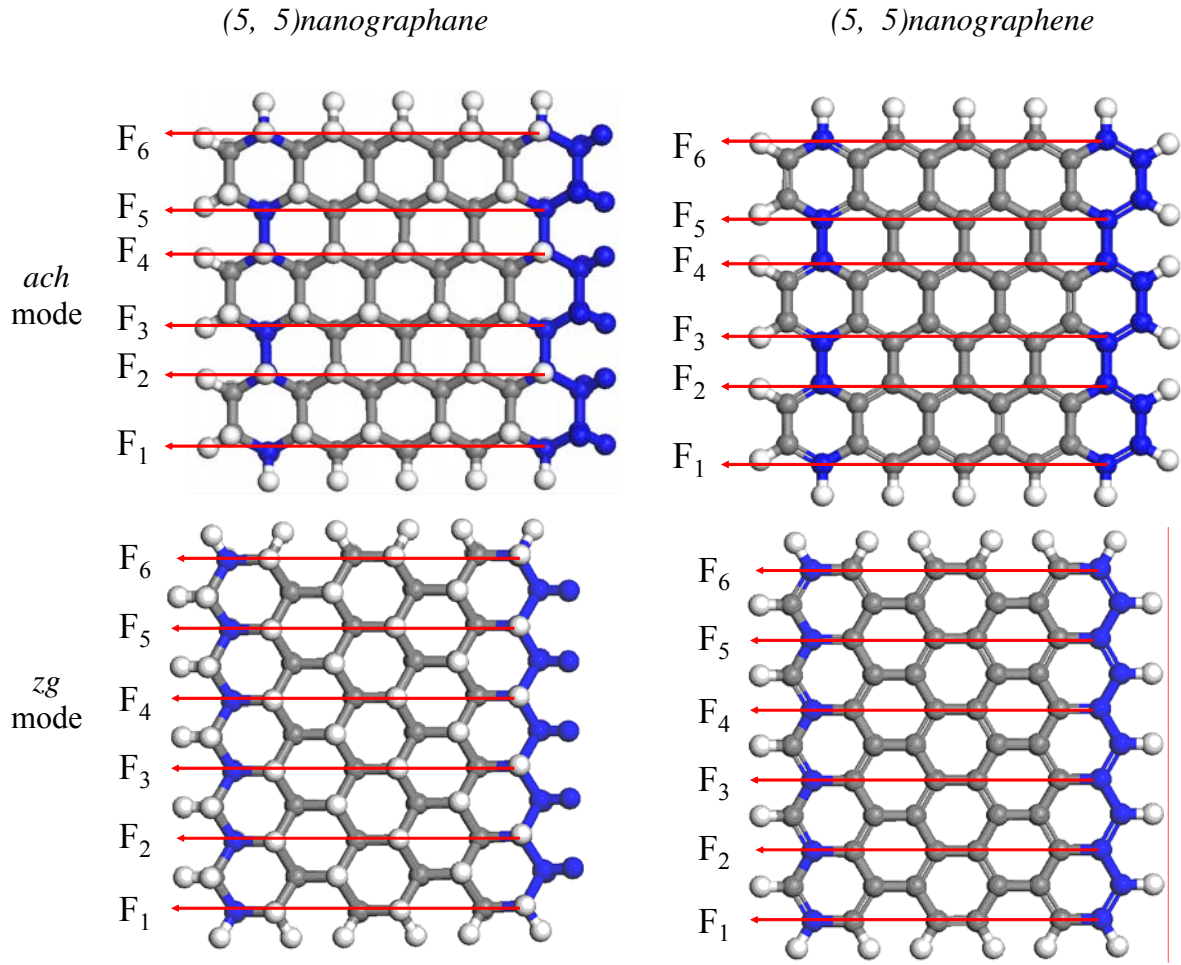

**Figure 1**. Configuration of six MICs related to two deformation modes of graphane and graphene (5,5) nanosheet. White and gray balls mark hydrogen and carbon atoms in equilibrium position. Edge carbon atoms are terminated by two and one hydrogen atoms for graphane and graphene, respectively. Atoms marked in blue are excluded from the optimization procedure. We have used the $(n_a, n_z)$ nomenclature of rectangular nanographenes suggested in [19].

Specifying uniaxial tension as a particular deformation mode within the framework of the QMRC approach [18], one has to introduce specific mechanochemical internal coordinates (MICs). The MICs related to graphane in the current study were aligned either along the C-C bonds or normally to the latter as shown in Fig.1. The two sets distinguish two deformational modes that correspond to tensile deformation applied to zigzag and armchair edges of the sheets due to which they are nominated as *zg* and *ach* modes as was done previously in the case of graphene [11, 12]. The right ends of the MICs in Fig.1 are clamped (the corresponding carbon atoms are immobilized) while the left ends move in a stepwise manner with increment $\delta L=0.1$Å at each step so that the current MIC length constitutes $L = L_0 + m\delta L$, where $L_0$ is the initial length of the MIC and $m$ counts the number of the deformation steps. According to the QMRC concept, the MICs are excluded from the optimisation procedure when seeking the minimum of the total energy. A *force of response* is determined as the residual gradient of the total energy along the selected MIC. These partial forces $F_i$ are used afterwards for determining micro-macroscopic mechanical characteristics related to uniaxial tension, among which there are:

- Force of response $F$:

$$F = \sum_i^{MIC\_number} F_i ; \quad (1)$$

- Stress $\sigma$:

$$\sigma = F/S = \left(\sum_i^{MIC\_number} F_i\right)/S, \quad (2)$$

where $S$ is the loading area determined as $S = DL_{0\_z(a)}$ and where $D$ is either the van der Waals thickness of the $sp^3$-bending carbon-atom chain of 4.28Å (graphane) or the van der Waals diameter of carbon atom of 3.34Å (graphene) and $L_{0\_z(a)}$ is the initial length of the MICs in the case of *ach* and *zg* modes, respectively;

- Young's modulus $E$:
$$E = \sigma/\varepsilon, \quad (3)$$

where both stress $\sigma$ and strain $\varepsilon$ are determined within the elastic region of deformation that starts at the MIC's length $l_0$ so that

$$\varepsilon = (L_i - l_0)/l_0 \quad (4)$$

in the general case [18]. The current length $L_i$ is related to a particular deformation mode and within the mode is identical for all the MICs.

The QMRC concept was implemented in the DYQUAMECH software [20] used in the study. The program is based on the Hartree-Fock unrestricted version of the CLUSTER-Z1 codes exploiting advanced semiempirical QCh method (PM3 in the current study) [21]. The program retains all features of the unrestricted broken symmetry approach, particularly important for odd electronic systems of graphene [22]. A completed computational cycle provides the following data.

*Microscopic characteristics* that include
(i) atomic structure of the loaded sheet body at any stage of the deformation including C-C bond scission and post-breaking relaxation;
(ii) strain energy

$$E_S(\varepsilon) = E_{tot}(\varepsilon) - E_{tot}(0) \quad (5)$$

where $E_{tot}(0)$ and $E_{tot}(\varepsilon)$ are the total energy of unloaded sample and sample subjected to $\varepsilon$ strain, respectively;
(iii) both partial $F_i$ and total $F$ (1) force of response;
(iv) molecular and atomic chemical susceptibilities of the body expressed in terms of the total $N_D$ and atomically partitioned $N_{DA}$ numbers of effectively unpaired electrons [23], respectively.

The dependences of energy, force, stress and $N_D$ on either elongation or strain exhibit mechanical behaviour of the object at all the stages of the deformation considered at the atomic level [11, 12]. The latter presents changing in the chemical activity of the body in due course of deformation.

*Micro-macroscopic characteristics* involve stress-strain interrelations (2), a forthcoming analysis of which allows for determining areas related to elastic deformation thus providing grounds for the elasticity theory application aimed at obtaining elastic mechanic parameters.

**3. Nanographene and/or nanographane under tensile deformation**

3.1. *ach* Mode of the deformation

The identity of the carbon cores of both pristine graphene [11, 12] and graphane (5,5) nanosheets makes it possible to skip some details concerning the description of the sheet mechanical behavior presented in [11, 12] and to mainly concentrate on this behavior difference under hydrogenation. A sketch of the main characteristics related to the *ach* mode of the graphene tensile deformation and failure is shown in Fig.2a.

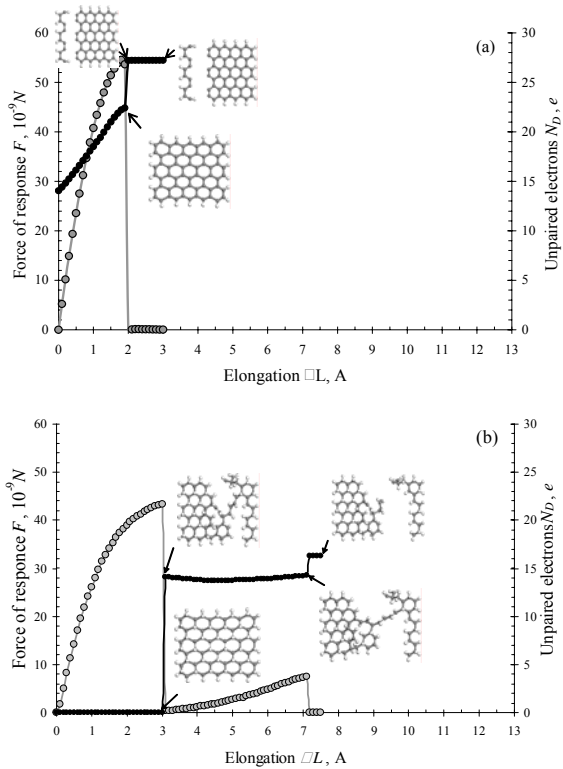

**Figure 2**. *ach* Deformation mode. Force-elongation (gray curves with dots, left Y-axis) and $N_D$-elongation (black curves with dots, right Y-axis) dependences and selected sets of structures related to (5,5) nanographene (a) and nanographane (b) (see text). Arrows point the attribution of the structures to the dots on the $N_D$-elongation curves.

As discussed in [11, 12], the sheet is uniformly stretched in due course of first 19 steps and breaking the C-C bonds occurs at the 20$^{th}$ step [24]. The sheet is divided in two fragments, one of which is a shortened (5,4) equilibrated nanographene while the other presents an alternative acetylene-and-carbene chain. The force-elongation dependence shows one-stage process of the sheet deformation and failure. The $N_D$-elongation exhibits a gradual increase in the sheet chemical activity up to the 19$^{th}$ step caused by a gradual elongation of C-C bonds, after which an abrupt grow of the chemical activity occurs at the 20$^{th}$ step caused by a simultaneous rupture of six C-C bonds. The achieved $N_D$ value remains then constant and not dependent on the distance between two fragments of the broken nanographene. Structures presented in the figure are related to the 19$^{th}$, 20$^{th}$ and 30$^{th}$ steps of the reaction.

The *ach* deformation mode of graphane, as seen in Fig.2b, occurs quite differently. The force-elongation curve evidences a two-stage deformation process. During the first stage, the nanosheet is firstly uniformly stretched similarly to graphene, but the stretching area is enlarged up to the 31$^{st}$ step. The stretching occurs without any C-C bond breaking so that, as it should be expected for the *sp³* electronic system without odd electrons, the number of effectively unpaired electrons $N_D$ remains zero up to the 30$^{th}$ step. At the 31$^{st}$ step $N_D$ abruptly grows up to 14 *e*, thus making the deformed sheet highly chemically active. The $N_D$ growth is caused by the rupture of a number of C-C bonds and the formation of alkene linear chain that connects two fragments of the sheet. Additionally, a small CH-unit cluster is formed in the upper part of fragment 2. Further elongation causes a straightening of the chain, which takes additional 40 steps, after which the chain breaks and two separate fragments are formed. The chain breaking is accompanied by the second sharp $N_D$ grow and the achieved value remains unchanged when further elongation proceeds. Structures presented in the figure are related to the 30$^{th}$, 31$^{st}$, 70$^{th}$, and 71$^{st}$ steps of the reaction. Comparing results presented on panels *a* and *b* of the figure, one can see 1.5 enlarging of the deformation area at the first step of deformation in graphane.

The *ach* deformation mode of graphane, which similarly to the *zg* deformation mode of graphene is accompanied with the formation of an extended carbon chain, alkene chain in the current case, is obviously size-dependent since the larger sheet the longer chain will be produced and the bigger number of steps will be required to stretch and break the chain. Analyzing the structure of the upper part of the sheet in Fig. 3b releases four CH groups that completed the cyclohexanoid unit in the non-deformed body but compose $(CH)_4$ cluster in the upper right angle of the structure when the alkene chain forms. When applying this algorithm to larger (15, 12) nanosheet shown in Fig.3c, a clearly seen trajectory of the alkene chain formation can be distinguished. As previously, four carbon atoms form a fastener in the bottom part of the structure while four other carbon atoms form a cluster in the upper side. The scheme in Fig. 3c may give an idea about the dependence of the deformation process on the sheet size.

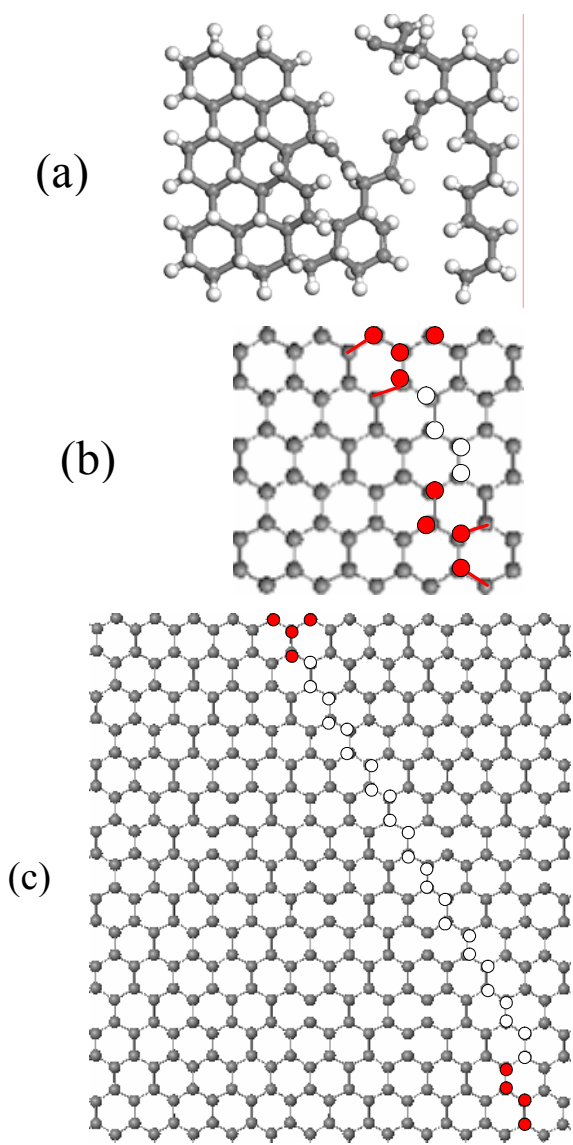

3.2. *zg* Mode of the deformation.

Figure 4a presents of a concise set of structure images of successive deformation steps revealing an exciting picture of a peculiar 'tricotage-like' failure of the pristine graphene body [11, 12]. In terms of this analogy, each benzenoid unit presents a stitch. In the case of the *ach* mode shown in Fig.2a, the sheet rupture has both commenced and completed by the rupture of a single stitch row. In the case of the *zg* mode, the rupture of one stitch is 'tugging at thread' the other stitches that are replaced by still elongated one-atom chain of carbon atoms. This causes a saw-tooth pattern of both force-elongation and $N_D$ -elongation dependences. In contrast to the multi-stage behavior of the pristine graphene, the *zg* mode of the graphane failure has commenced and completed within the first stage (see Fig.4b). Similarly to the situation occurred for the *ach* mode, the critical elongation in the case of graphane is ~1.5 times bigger that one related to the first stage of the graphene deformation.

Looking at the data presented in Figs. 2 and 4, one can conclude that mechanical behavior of graphane differ from that one of graphene quite significantly. First, the hydrogenation effectively suppresses the tricotage-like pattern of the graphene deformation and failure. The hydrogen effect has been firstly shown in [11, 12] when comparing the deformation of the graphene bare-edge nanosheet with that deformed (5, 5) sheet at the 31$^{st}$ step, it is possible to suggest a probable algorithm of the alkene chain formation that is presented in Fig. 3. As can be seen in Fig. 3a, the rupture starts in the right angle of the sheet bottom by breaking two C-C bonds marked by red in Fig. 3b. Adjacent four red atoms form the chain fastener while the chain itself is formed by white atoms along alkyl configuration of carbon atoms in the pristine graphane. Breaking two C-C bonds in the one related to the sheet with hydrogen-terminated edges: this latter case is presented in Figs. 2 and 4. Sixteen-tooth force-elongation and $N_D$-elongation dependences of bare-edge sheet are shorten to the eight-tooth pattern after termination of the edges by hydrogen atoms. However, once suppressed, the tricotage-like behavior has been still well exhibited in this case as seen in Fig.4a. A further hydrogenation, which concerns carbon atoms on the sheet basal plane, smoothes the behavior further transferring the eight-tooth dependences in Fig.4a into two-tooth ones in Fig.2b thus shortening the scale of the deformation area from 12.1 to 7.2Å. Second, the multi-tooth pattern is characteristic for the *zg* mode of both bare-edge and hydrogen-terminated-edge of pristine graphene [11, 12] in contrast to one-stage deformation related to the *ach* mode thus showing the sheet to be more deformable along C-C bonds (see Fig.1).

**Figure 3**. A schematic presentation of the origin of alkene chain in the graphane body at the *ach* mode of deformation. a. Equilibrium structure of the (5, 5) graphane sheet at the 31$^{st}$ step of deformation. b and c. Schemes of the chain formation in (5, 5) and (15, 12) nanosheets. Hydrogen atoms are not shown.

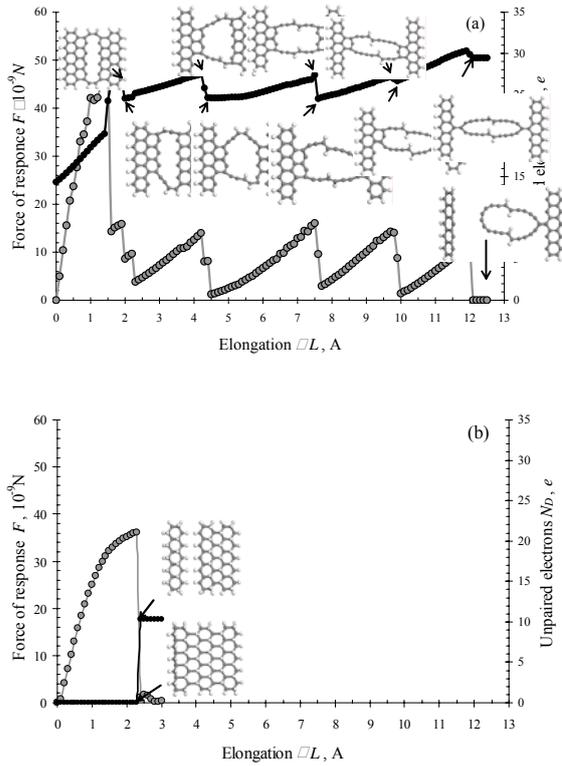

**Figure 4**. *zg* Deformation mode. Force-elongation (gray curves with dots, left Y-axis) and $N_D$-elongation (black curves with dots, right Y-axis) dependences and selected sets of structures related to (5,5) nanographene (a) and nanographane (b) (see text). Arrows point the attribution of the structures to the dots on the $N_D$-elongation curves.

Oppositely, a saw-tooth pattern is observed for the *ach* mode of the graphane deformation exhibiting higher deformability of the sheet along a C-C-C alkyl chain. The difference in the mechanical behavior related to *ach* and *zg* modes of graphene was attributed to the different configurations of the benzenoid stitch packing with respect to the body C-C bond chains. The same reason seems to lay the foundation of the anisotropy observed for the graphane sheet but it should be addressed to the cyclohexanoid units packing.

## 4. Energetic characteristic of the graphane tensile deformation

The total energy of the graphane nanosheet subjected to tensile stress is presented in Fig.5a. At first glance, the plotting looks quite typical for condensed media so that the application of the linear elasticity approach in the region close to the initial MIC's length seems to be quite possible. However, more detail plotting of forces of response in Fig.5b highlights some inconsistence with the approach in this region. As seen in the figure, the first two points on both curves do not subordinate the linear dependence and match the region the rubbery high-elasticity state introduced for polymers where a similar situation has been met quite often [26, 18]. Short plots of linear dependences are located between the fourth and the eighth points. Approximating them to the intersection with the abscissa allows for obtaining reference MIC lengths $l_0$ that correspond to the start of the linear dependence and match the beginning of the elastic deformation. Thus determined $l_0$ values are taken into account when determining strain ε expressed by (4) to plot the strain energy $E_s(\varepsilon)$-strain dependences in Fig.5c, as well as stress-strain dependences in Fig. 6. Both plottings are used when determining Young's modules for the considered deformation modes.

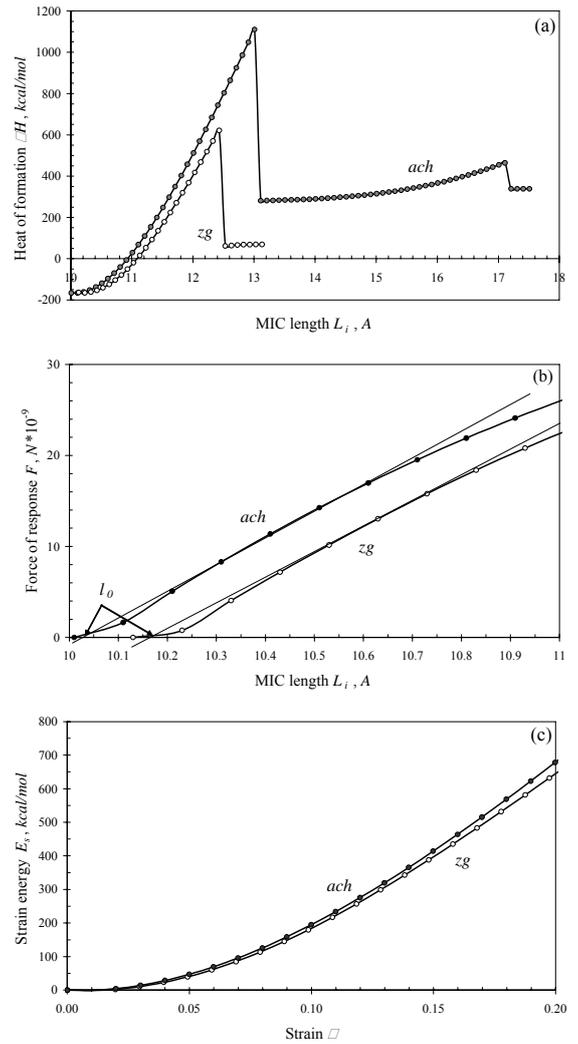

**Figure 5**. Dependences of total energy (a) and force of response (b) on MIC's length as well as strain energy on strain (c) of (5,5) nanographane. Filled and empty circles match *ach* and *zg* deformation modes, respectively.

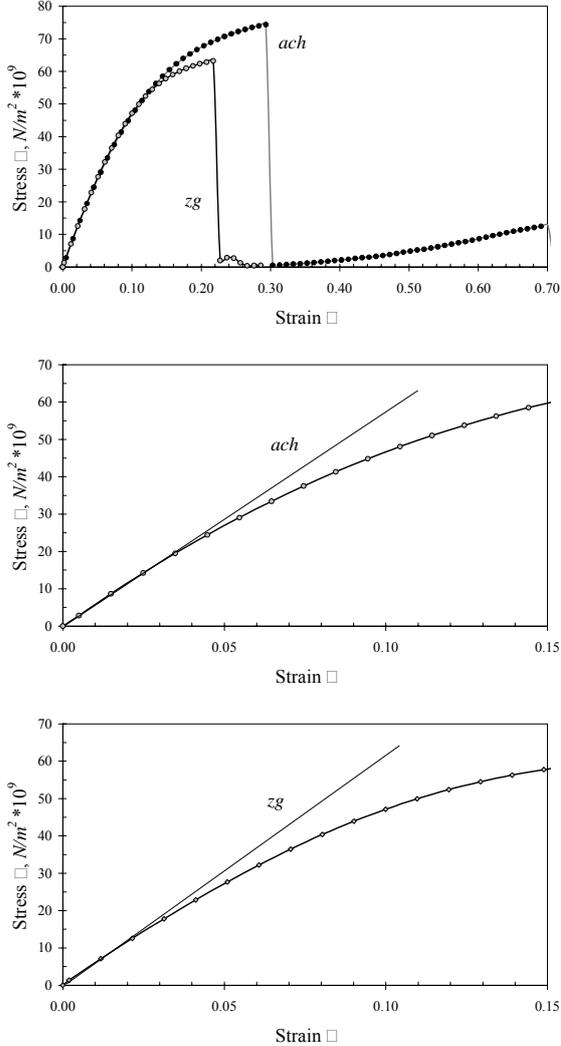

**Figure 6**. Stress-strain relationships for (5,5) nanographane subjected to tensile deformation (a) and that ones at the first deformation steps related to *ach* (b) and *zg* (c) modes.

Figure 6a highlights the difference in the mechanical behavior of *zg* and *ach* deformation modes of graphane. Leaving aside the two-tooth pattern of the stress-strain dependence for the *ach* mode, one can concentrate on the first stages of the deformation in both cases that evidently govern quantitative values of the deformation parameters. As seen in the figure, both modes are practically non-distinguishable up to 12% strain after which the failure process proceeds much quicker for the *zg* mode and accomplishes at the 23% strain while the failure related to the *ach* mode continues up to 30% strain. The critical values of force of response $F_{cr}$, stress $\sigma_{cr}$, and strain $\varepsilon_{cr}$ are presented in Table 1.

Once non-distinguishable by eyes, the stress-strain regularities of the two modes related to the first stage of the deformation are nevertheless different as shown in Figs. 6b and 6c. Firstly, the difference concerns the size of the region attributed to the elastic behavior of the plottings. As seen in the figures, the *ach* deformation is elastic up to the strain reaching 4% while in the *zg* mode the elastic region is shorter and is commenced on 2% strain. However, practically about half of the region in both cases is not elastic but corresponds to the rubbery high-elasticity state as shown in Fig.5b. Therefore, the graphane deformation is mainly non-elastic, once rubbery high-elastic at the very beginning and plastic after reaching 4% (*ach*) and 2% (*zg*) strain. The values of Young's moduli $E_\sigma$ attributed to the elastic parts of stress-strain plottings in Figs. 6b and 6c are given in Table 1.

As well known, another way to determine Young's modulus is addressed to the strain energy (5) that in the elastic approach can be presented as [26]

$$E_s(\varepsilon) = 1/2 v_0 \mathrm{E}_e \varepsilon^2, \qquad (6)$$

where $v_0$ is the volume involved in the deformation and $\mathrm{E}_e$ is the related Young modulus. Applying the approach to the initial part of $E_s(\varepsilon)$ curves presented in Fig.5c, it has been revealed that the elastic approach fits the curves at the same regions of strains as determined above for the stresses-strain curves. Obtained $\mathrm{E}_e$ values in these regions are well consistent with the previously determined $E_\sigma$, as seen from Table 1.

To conclude the discussion of peculiarities of mechanic characteristics of graphane, a comparative view on the mechanical behavior of graphene and graphane is presented in Fig.7. The presentation is limited to the first stage of the graphane deformation. The figure exhibits quite clearly

- The existence of the rubbery high-elastic state at the beginning of the deformation in graphane while a similar state in the graphene deformation is absent so that the deformation of the latter in this region is elastic and the origin of the elastic behavior $l_0$ coincides with the initial MIC's length $L_{0\_z(a)}$;
- Much less elasticity of graphane if one applies the elastic approach to the latter;
- Plastic regime of the main part of the deformation of both bodies;
- Termination of graphene failure at much shorter elongation than that of graphane.

Alongside with the peculiarities caused by the hydrogenation of the pristine graphene and discussed earlier that concern a suppress of the tricotage-like character of the deformation, all the findings suggest a quite extended picture of the hydrogenation effect on mechanical properties of graphene. A native question arises what reasons are behind the observed features. We suppose to find the answer addressing a molecular aspect of the deformation process.

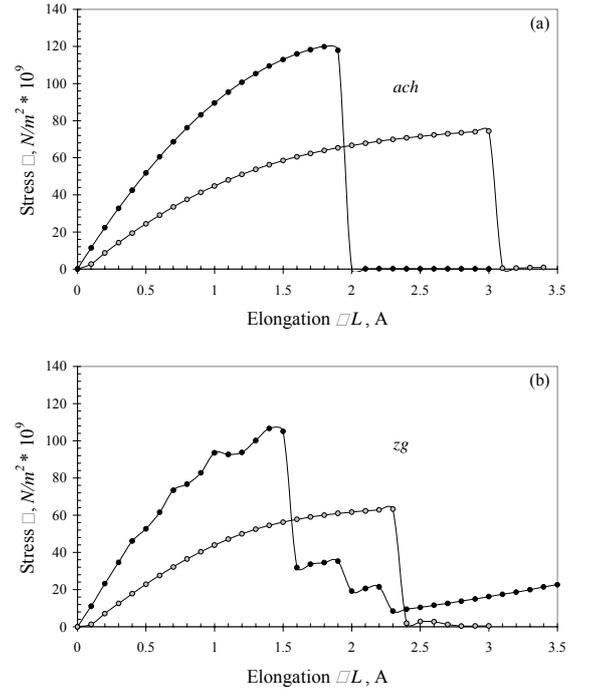

**Figure 7**. Stress-strain relationships for graphene (black dots) and graphane (gray dots) (5,5) nanosheets subjected to tensile deformation related to *ach* (a) and *zg* (b) modes at the first stages of deformation.

## 5. Molecular aspect of the graphene and/or graphane deformation

As was shown [11, 12], unique mechanic properties of graphene have originated from the mechanic resistance of its benzenoid units. Taking benzene molecule as their molecular image, the QMRC approach has revealed a vividly seen mechanical anisotropy of the molecule which lays the foundation of the difference in the *ach*- and *zg*-mode's behavior of graphene. Table 1 lists the main mechanical parameters of the molecule that are well consistent with those related to the graphene body thus demonstrating the origin of the body's properties. In this view it is natural to draw attention to the mechanical properties of a

*Tensile deformation of benzene*

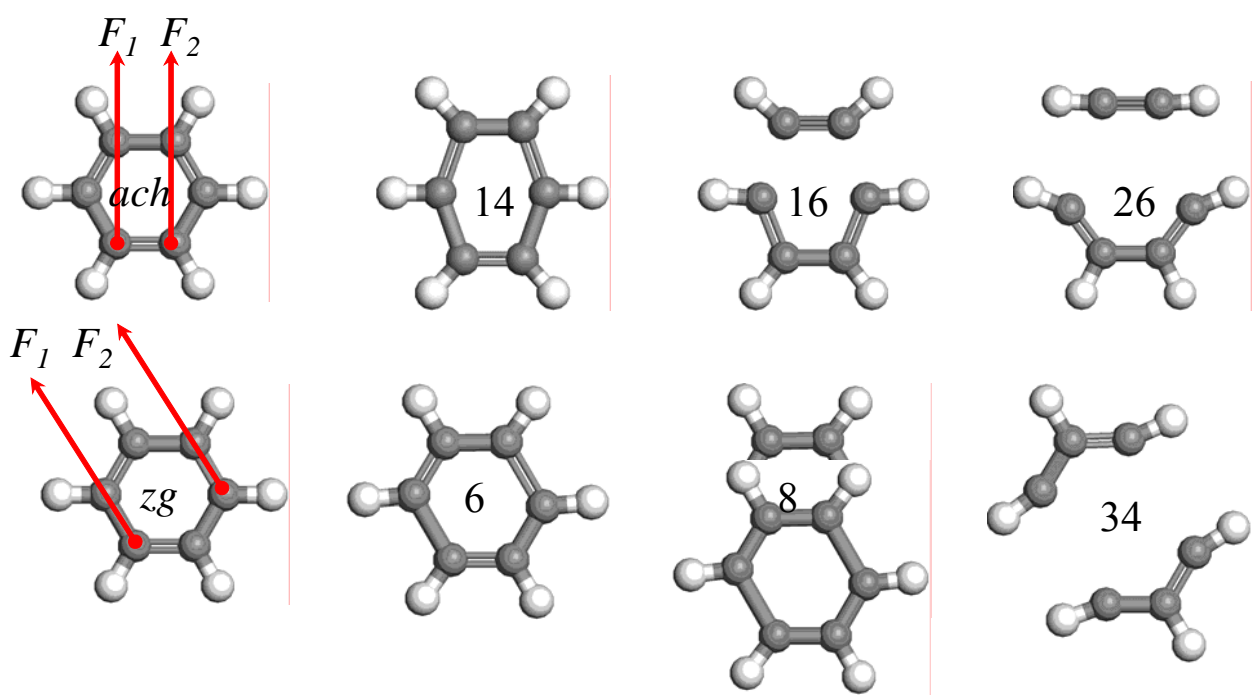

*Tensile deformation of cyclohexane*

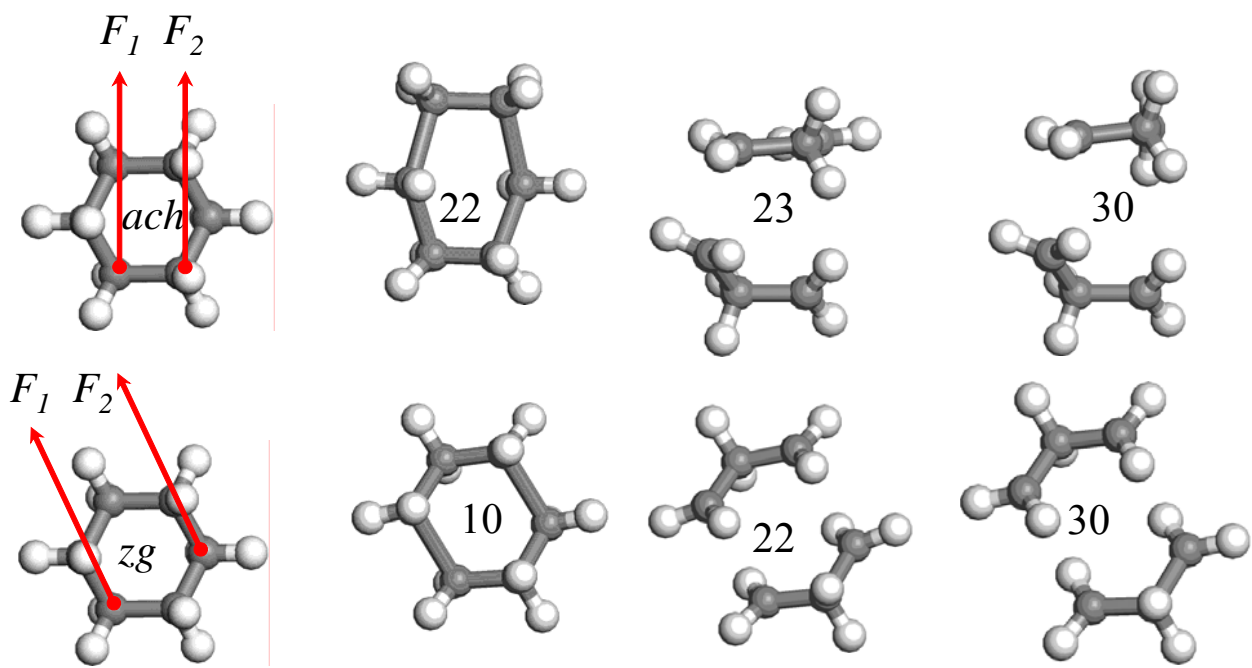

**Figure 8**. Equilibrium structures of benzene and cyclohexane molecules subjected to the stepwise tensile deformation. The configuration of MICs related to *ach* and *zg* modes are shown in the left panels. Figures point step numbers.

molecule that might simulate the cyclohexanoid unit of graphane. Obviously, cyclohexane in its chairlike conformation should be the appropriate model. Figure 8 presents a set of structures of the benzene and cyclohexane molecules subjected to uniaxial tensile deformation described by two pairs of MICs that can be attributed to the *ach* and *zg* modes discussed in the previous sections. Deformation proceeds as a stepwise elongation of both MICs at each deformation mode with a constant increment of 0.05Å per a step. Details of the computational procedure are given in [11, 12] for benzene molecule.

A comparative insight on the quantitative characteristics of the mechanical behavior of both molecules is summarized in Fig. 9. Stress-strain curves accumulate the mechanical behavior while $N_D$-strain curves exhibit changes in the chemical reactivity of the molecules in due course of the deformation. As seen in the figure, stress-strain curves show that sharp mechanic anisotropy of benzene is significantly smoothed for cyclohexane. The latter is evidently less elastic and is broken at less stress but at bigger strain in both modes. Changing in the chemical reactivity of the two molecules is evidently different. Owing to $sp^2$ character of the electron system of benzene, the $N_D$ value, equal to zero in the non-loaded state due to fitting the length of its C-C bond the critical value $R_{crit} = 1.395$ Å [23], gradually increases over zero in due course of the deformation due to lengthening of C-C bonds up to ε~0.4 and then abruptly grows due to the bond rupture. In contrast, $N_D$ value for $sp^3$ hybridized cyclohexane retains zero until the C-C bonds rupture occurs at 23$^{rd}$ and 22$^{nd}$ steps of the *ach* and *zg* modes, respectively, when an abrupt $N_D$ growth up to ~4 *e* occurs. It should be noted that if the C-C bonds rupture in cyclohexane at *ach* mode is accompanied by the immediate $N_D$ growth, in the case of *zg* mode the $N_D$ growth is somewhat postponed and occurs in 12 steps later until the formed fragments take the form of step 23 shown in Fig.8. The preceding 12 steps correspond to overstretched C-C bonds such as in the form of step 22 in Fig.8.

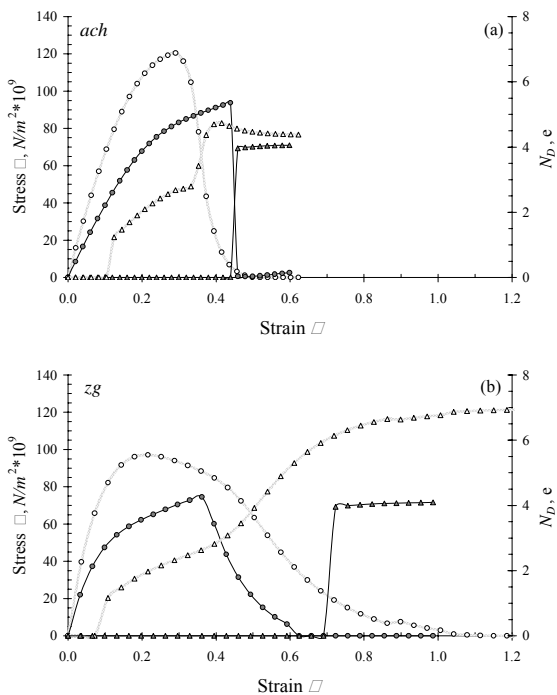

**Figure 9**. Stress-strain (circles, left Y-axis) and $N_D$-strain (triangles, right Y-axis) dependences related to benzene (empty) and cyclohexane (filled) at *ach* (a) and *zg* (b) deformation modes.

Mechanical characteristics of both molecules are presented in Table 1 alongside with those related to (5,5) nanographene and nanographane discussed in the previous sections. As clearly seen from the table, there is a close resemblance of the data related to the molecules and the condensed bodies. The very molecular characteristics explain why the graphane elasticity is lower, why the critical strains are bigger but critical stresses are smaller. Taking together these characteristics provide the lowering of Young's modulus of graphane in both deformation modes. Obviously, a complete coincidence of the molecular and condensed body data should not be expected since packing of both benzenoid and cyclohexanoid units must remarkably contribute into the mechanical behavior. The packing, which is anisotropic in regards C-C bond chains, seems to be responsible for the difference in the topology of the failure during the *ach* and *zg* modes. But, the asymmetry of the final products at the *ach* deformation of cyclohexane seems to be quite suitable for the creation of CH-unit chain when adjacent cyclohexanoid units are cracked.

## 6. Mechanical deformation and dynamic properties

Since every mechanical deformation is vibration-involving [26], there might be a tight connection between the mechanical behavior of the object and its vibrational and/or phonon spectrum. As shown for polymers [26], the elastic part of the tensile behavior, which is provided by stretching of chemical bonds, is obviously connected with the relevant valence vibrations. The occurrence of the rubbery highly-elasticity state is usually attributed to a high conformational ability of polymers that is provided by low-frequency torsional and deformational vibrations [18]. In view of this concept, elastic parameters such as Young's modulus and stiffness must by proportional to force constants, or squared frequency, of the relevant harmonic vibrations.

In the current case, the deformation of both benzene and cyclohexane molecules and graphene/graphane bodies are determined by the rupture of C-C bonds. Frequency of the characteristic C-C valence vibrations of benzene constitutes $v_{C-C}^{BZN} = 1599$ cm$^{-1}$ [27]. Taking $v_{C-C}^{BZN}$ as the reference, it is possible to determine the relevant frequency of cyclohexane $v_{C-C}^{CHXN}$ that has to provide the lowering of its Young's modules using a relation

$$v_{C-C}^{CHXN} = \eta v_{C-C}^{BZN}, \qquad (7)$$

where

$$\eta = \sqrt{\left(E_\sigma^{CHXN} / E_\sigma^{BZN}\right)}. \qquad (8)$$

Here $E_\sigma^{CHXN}$ and $E_\sigma^{BZN}$ are Young's modules listed in Table 1. The ratio of the Young's modules $E_\sigma^{CHXN} / E_\sigma^{BZN}$ alongside with η values and obtained $v_{C-C}^{CHXN}$ values are given in Table 2. As seen from the table, the latter are in good consistence with two molecular frequencies, related to stretching C-C vibrations of cyclohexane, computationally predicted in [28]. The vibration's eigenvectors point to significant difference in the vibrations shape, which may explain their selective involving into elastic deformation of cyclohexane at two different deformation modes. Column η$_{exd}$ presents expected η values obtained according to (7) by using calculated vibrational frequencies from column $v_{C-C}$.

Calculated phonon spectra of both graphene [29, 30] and graphane [31] are formed by broad bands whose attribution to C-C bond stretching is not so straightforward. In the case of graphene, a definite eigenvector phonon mode, which points to C-C stretching, is related to the optical phonons at the Γ point of the Brillouin zone that produces a characteristic G-band at 1564 cm$^{-1}$ observed in the species Raman spectrum [29]. Taking the frequency as the reference $v_{C-C}^{graphene}$, one can evaluate by using relation (8) the corresponding graphane frequencies $v_{C-C}^{graphane}$ that have to provide the observed decreasing of Young's modules presented in Table 2. The corresponding frequencies lie in 1095-1170 cm$^{-1}$ region. The calculated graphane phonon spectrum [31] exhibit pure carbon-atom-involved and mixed carbon-atom-involved internal phonons in the region from 1000 to 1330 cm$^{-1}$ thus determining the region for expected η$_{exd}$ value from 0.64 to 0.85, which is in a perfect consistence with the data presented in Table 2. Therefore, the lowering of Young's modulus of graphane with respect to graphene has a dynamic nature and is connected with softening of C-C internal phonons when passing from $sp^2$ to $sp^3$ configuration. The softening is caused by weakening the corresponding C-C bonds, which was experimentally proven just recently [32].

Besides valence vibrations, hydrogenation of both benzene molecule and graphene provides the appearance of low-frequency torsional and deformational vibrations [28, 31]. These vibrations do not influence much the deformation of an individual molecule but may affect significantly the behavior of condensed chains of the latter. It seems quite reasonable to suggest that the rubbery high-elastic state of the graphane deformation is just a highlighting of a considerable softening of this low-frequency part of the phonon spectrum of graphane [31] with respect to graphene.

The dynamic origin of the Young modulus lowering has forced us to draw attention on the elastic parameters of fluorographene (CF)n [33] whose phonon spectrum has been calculated [31]. According to calculations, C-C internal phonons fill the region of 1100-1330 cm$^{-1}$, due to which one can expect decreasing of the Young modulus for the fluoride from 0.49 to 0.72, which is similar to graphane. The first observations [33] have shown the Young's modulus of fluorographene constituting ~0.3 that of graphene. Once complicated by difficulties inherited in the fluorographene production and indentation technique, the data are in a reasonable consistence with the prediction.

## 7. Conclusion

In summary, we have carried out systematic quantum-mechanochemical-reaction-coordinate simulations to study the mechanical properties of graphane. The latter implies the chairlike conformer of cyclohexanoid units of the honeycomb packing of carbon atoms. We have found that the mechanical behavior of graphane is anisotropic similarly to graphene so that tensile deformation along (*zg* mode) and normal (*ach* mode) C-C bonds chain occurs quite differently. Nevertheless, typical to graphene tricotage-like pattern of the body failure is considerably suppressed. Besides that, the spatial region of elastic regime is remarkably shortened and a rubbery high-elasticity state has been exhibited. The tensile strengths at fracture constitute 62% and 59% of graphene for *ach* and *zg* modes, respectively while the fracture strains increase therewith by 1.7 and 1.6 times. Young's modules of the two deformation modes of graphane decrease by 1.8 and 2 times. Our simulations show that the main part of the change occurs at a molecular level and is provided by the substitution of benzenoid units of graphene by cyclohexanoid ones of graphane. Additional contribution is connected with the difference of the units packing in the honeycomb bodies. A vibration-involving deformation concept has been successfully applied to explain the decrease of the Young modules connecting the latter with softening C-C stretching vibrations in due course of the $sp^2$-to-$sp^3$ bonding transition. A justified prediction concerning elastic parameters of fluorographene has been made.

**Acknowledgement**
The authors are grateful to L.Kh.Shaymardanova for assistance with modeling related to benzene and cyclohexane molecules.

**Table 1**. Mechanical parameters of benzene and cyclohexane molecules as well as graphene and graphane nanosheets

| Species | Mode | $\varepsilon_{cr}$ | $F_{cr}$, N·$10^{-9}$ | $\sigma_{cr}$, N/m²·$10^9$ | $E_{\sigma,e}$, TPa |
|---|---|---|---|---|---|
| benzene | *ach* | 0.29 | 19.62 | 120.47 | 0.76 |
| | *zg* | 0.22 | 19.18 | 97.13 | 0.99 |
| cyclohexane | *ach* | 0.44 | 15.69 | 93.76 | 0.4 |
| | *zg* | 0.36 | 14.99 | 74.57 | 0.74 |
| (5, 5) nanographene | *ach* | 0.18 | 54.56 | 119.85 | 1.09 |
| | *zg* | 0.14 | 47.99 | 106.66 | 1.15 |
| (5, 5) nanographane | *ach* | 0.3 | 43.41 | 74.37 | $0.61_\sigma$ ($0.54_e$) |
| | *zg* | 0.23 | 36.09 | 63.24 | $0.57_\sigma$ ($0.52_e$) |

**Table 2**. Young's modulus and frequencies of stretching C-C vibrations

| Species | Def. mode | $E_\sigma^{CHXN}/E_\sigma^{BZN}$ | $\eta$ | $\nu_{C-C}^{CHXN}$, cm⁻¹ | $\nu_{C-C}$, cm⁻¹ | $\eta_{exd}$ |
|---|---|---|---|---|---|---|
| BZN | ach | | | | 1599 [27] | |
| | zg | | | | | |
| CHXN | ach | 0.53 | 0.73 | 1160 | 1070 [28] | 0.67 |
| | zg | 0.75 | 0.86 | 1382 | 1388 [28] | 0.87 |
| | | | | $\nu_{C-C}^{graphane}$, cm⁻¹ | | |
| graphene | ach | | | | 1564 [29] | |
| | zg | | | | | |
| graphane | ach | 0.56 | 0.75 | 1170 | 1000-1330 [31] | 0.64-0.85 |
| | zg | 0.50 | 0.71 | 1095 | | |